\documentclass[%
 reprint,
 amsmath,amssymb,
 prd
]{revtex4-2}

\usepackage{physics}
\usepackage{graphicx}
\usepackage{color}
\usepackage{soul}

\begin{document}

\title{Ring of attraction: overlapping directions of the dipole modulation of the CMB, the parity asymmetry, and kinematic dipole percolation zone}

\author{James Creswell}
\affiliation{Niels Bohr Institute, University of Copenhagen, Blegdamsvej 17, DK-2100 Copenhagen, Denmark}

\author{Pavel Naselsky}
\affiliation{Niels Bohr Institute, University of Copenhagen, Blegdamsvej 17, DK-2100 Copenhagen, Denmark}

\begin{abstract}
The largest anisotropy in the cosmic microwave background (CMB) is the 3 mK kinematic dipole reflecting our motion with respect to the CMB frame and pointed in the direction $(l, b) = (264^\circ, +48^\circ)$ in Galactic coordinates.
We introduce the concept of the ring of attraction (RA), which is orthogonal to the axis of the kinematic dipole.
These directions overlap with the zone of percolation for the kinematic dipole, where its amplitude almost vanishes.
We show that along this ring are oriented the directions of the dipole modulation of the CMB, and positions of the peaks responsible for generation of parity asymmetry.
This coincidence is peculiar at greater than the 3 sigma level.
We analyzed the ``interaction'' of low multipoles of the CMB with RA and showed that for odd modes there is a sequence of peaks in the RA direction. 
These peaks correlate with each other for different multipoles and result in mutual amplification of the odd $\ell$ signal for the first 30 multipoles.
Our method sheds new light on the nature of parity asymmetry.
It consists of the deficit of symmetrically located and equal in amplitude peaks in the CMB map in comparison with asymmetric peaks.

\end{abstract}

\maketitle

\section{Introduction} \label{sec:intro}

The kinematic dipole is induced in the observed CMB due to the motion of the solar system with respect to the rest frame of the CMB radiation.
Observations from COBE, Planck and WMAP give a direction of the kinematic dipole $T_d(\mathbf{n})$ of $(l, b) = (264.00^\circ\pm 0.03^\circ, 48.24^\circ\pm 0.02^\circ)$ (Planck 2015 nominal) and $(l, b) = (263.99^\circ\pm 0.14^\circ, 48.26^\circ\pm 0.03^\circ)$ (WMAP) in Galactic coordinates $(l,b)$ \citep{Adam:2015rua,2011ApJS..192...14J}.
Denoting this direction with the unit vector $\mathbf{q}$, 
then 
we introduce the concept of the ring of attraction (RA) including all directions $\mathbf{g}$ such that
\begin{equation}
    \mathbf{q} \cdot \mathbf{g} = 0,
    \label{eq:dist}
\end{equation}
i.e.\ $\mathbf{g}$ are all unit vectors orthogonal to the direction of the kinematic dipole $\mathbf{q}$. These directions form  RA, which coincides with the percolation line of the kinematic dipole map, where $T_d(\mathbf{n}) = 0$. 

The purpose of our paper is to analyze some anomalies of temperature maps and the relationship of their characteristic directions in the sky both with each other and with RA. We restrict ourselves to an analysis of the  parity
asymmetry \citep{Kim_2010}, dipole modulation of the 
CMB temperature \citep{Eriksen_2004,Hansen_2009} and the coincidence of the directions of the quadrupole and octupole (quadrupole--octupole alignment \citep{Schwarz_2004}). These anomalies have been investigated in recent literature, where they are found to persist in the Planck data \cite{Schwarz:2015cma,Muir_2018,Shaikh_2019} (see also \citep{Vitenti:2019poh,Efstathiou:2020wem}).
 
We developed the theory of parity asymmetry in the pixel domain and
and showed that the dominance of the power of odd harmonics (asymmetric modes) over even (symmetric modes)
localized in the range $\ell = 20$ to $30$, in full accordance with the results of previous studies \citep{Creswell:2021eqi}. In the CMB map, this anomaly is associated with two pairs of high peaks (opposite in amplitudes) that belong to RA. The same directions are characteristic for the dipole modulation of CMB.

In this work, we investigate the peak structure of the $Z(\mathbf{n})=T(\mathbf{n})T(\mathbf{-n})$ asymmetry estimator.
The significance of the alignment of the peaks with the RA is estimated based on the pixel-domain distance and compared to Gaussian simulations.
This analysis confirms the significance at the level of around or better than 1 in 2000.

As part of our method, we reanalyze the problem of low multipoles, and showed that the peaks of $Z(\mathbf{n})$ for octupole ($\ell=3$) are oriented along of the RA. The same effect is typical for the quadrupole, unlike all other even $\ell$-modes. For the odd multipoles in the distribution of peaks for each $\ell>3$ we have found the sequence of
subdominant peaks adjusted to RA. The existence of such correlated sequences  between different odd $\ell$ peaks leads to formation of the high peaks in the CMB map with $\ell\le 20$ to $30$, which cross the RA at the direction close to the direction of the dipole modulation of CMB.

The outline of the paper is the following. In Section II we will present decomposition of the temperature map into symmetric and asymmetric modes and introduce the estimator $Z(\mathbf{n})$ of symmetry/asymmetry of signal for each pixel. We will present theoretical basis for distribution function $\mathcal{P}(Z)$  and for the parameter of asymmetry $R(\mathcal{P}(Z))$. We show that for  $2\le\ell \le 30$ the departure $R(\mathcal{P}(Z))$ from Gaussian statistic corresponds to 3$\sigma$ confidence level. Section III  is devoted to investigation of the common directions for the dipole modulation of the CMB and the parity asymmetry. We find that coincidence of 1a/1b peaks of $Z(\mathbf{n})$ and the direction of the dipole modulation  $(l,b)\approx(224^o,-22^o)\pm24^\circ$ occurs in 1/2000 cases  for Gaussian simulations. In Section IV we reanalysed the morphology of the low ($\ell=2$ to $7$) multipoles and showed that the RA is presented in the octupole at the dominant level, and for $\ell=5,7$ as a subdominant correlated sequences of peaks $Z(\mathbf{n})$.
The quadrupole, unlike other even $\ell$ modes, reveals the same tendency, which makes it abnormal. We summarized all the results in Conclusion.

\section{Symmetric and asymmetric parts of the temperature anisotropy}

\begin{figure}[!t]
    \centerline{
    \includegraphics[width=0.4\textwidth]{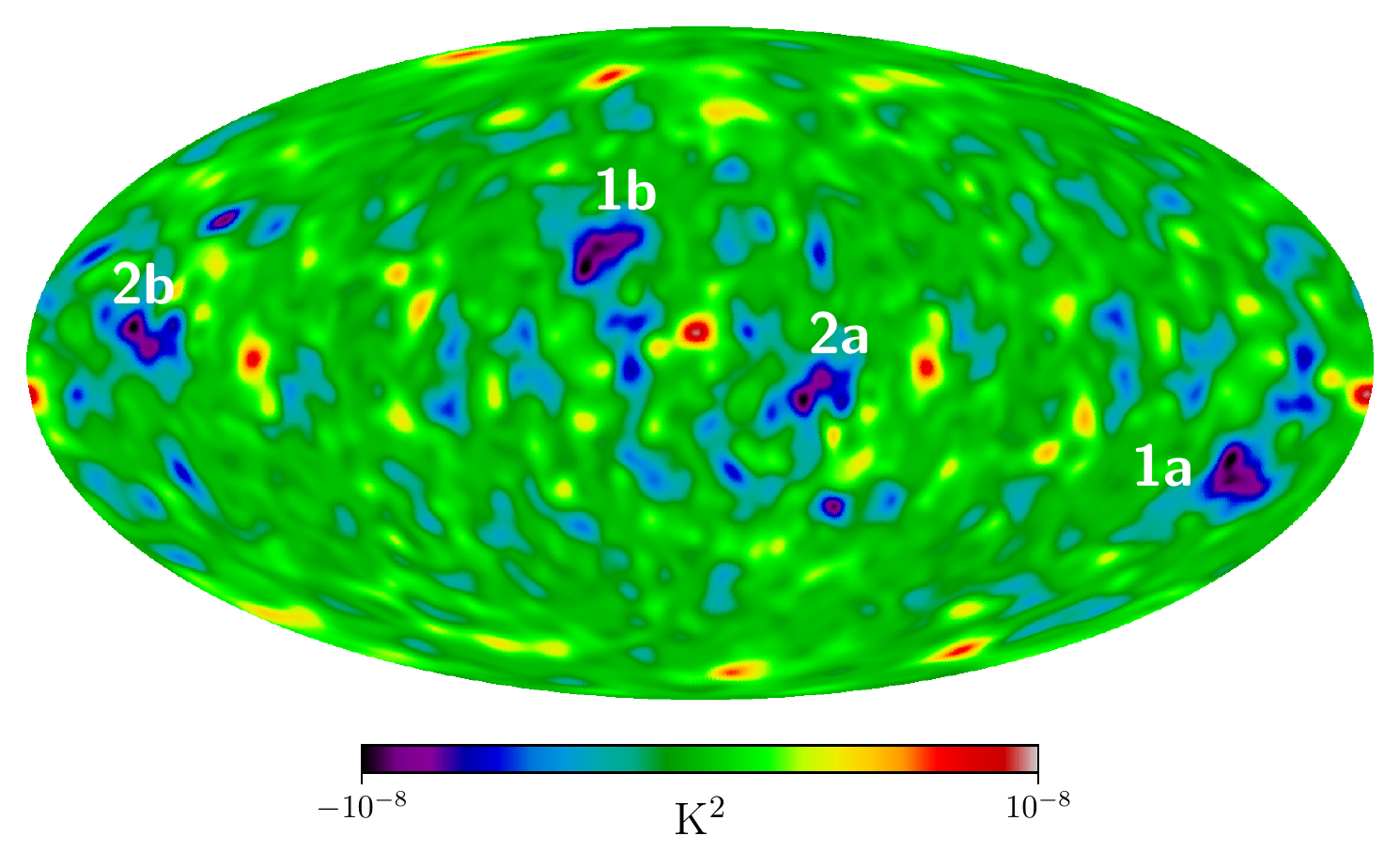}}
    \centerline{    
    \includegraphics[width=0.4\textwidth]{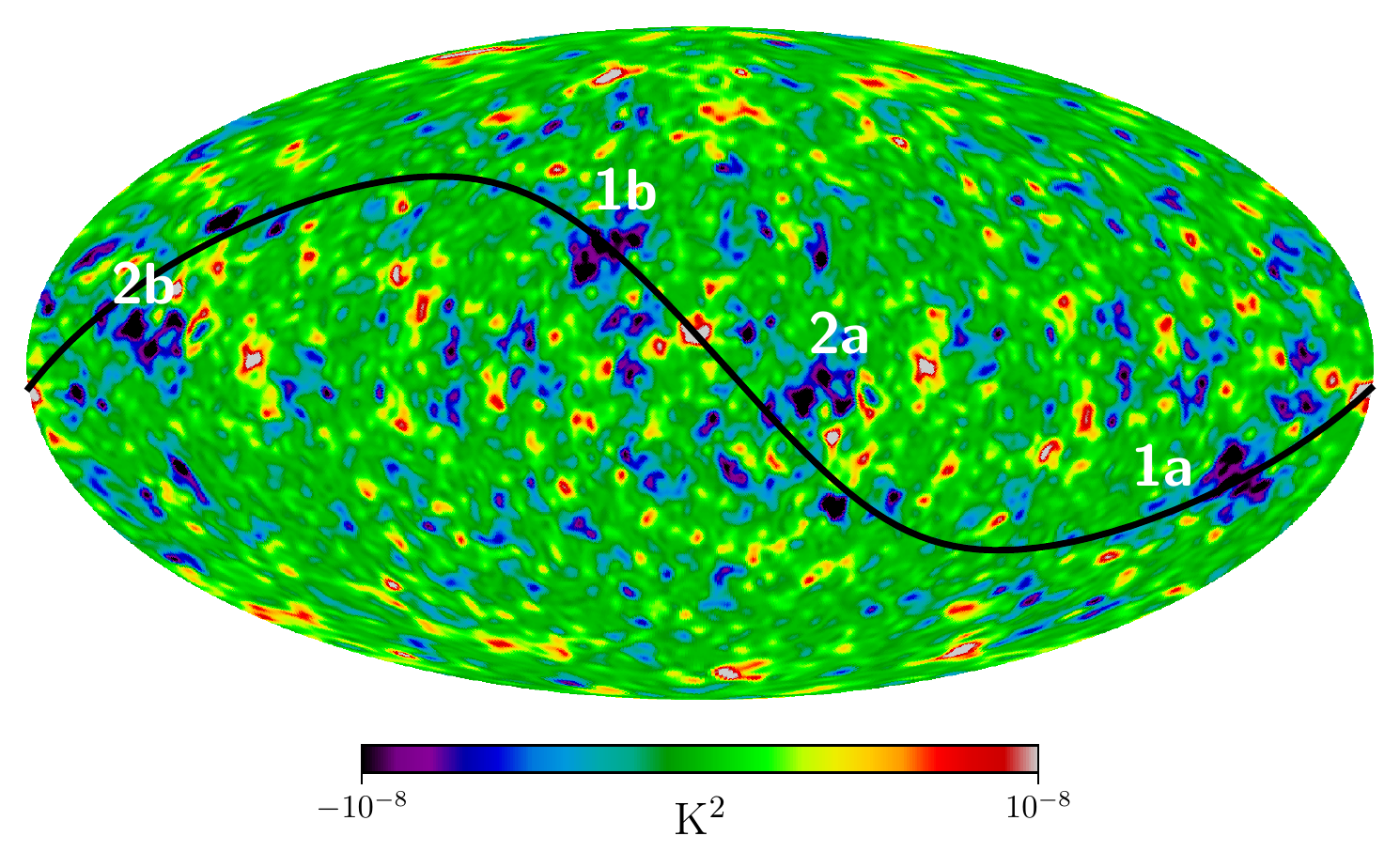}}   
    \caption{Top: $Z({\bf n})$ map of SMICA with $\Theta=5^\circ$ smoothing.
    Bottom: the same as top, but for $\Theta=2.5^\circ$. The black solid line
    indicates the ring of attraction, the direction on the sphere orthogonal to the direction of the kinematic dipole (the axis of evil).
    }
    \label{fig1}
\end{figure}

\begin{figure}[t]
    \centerline{
    \includegraphics[width=0.4\textwidth]{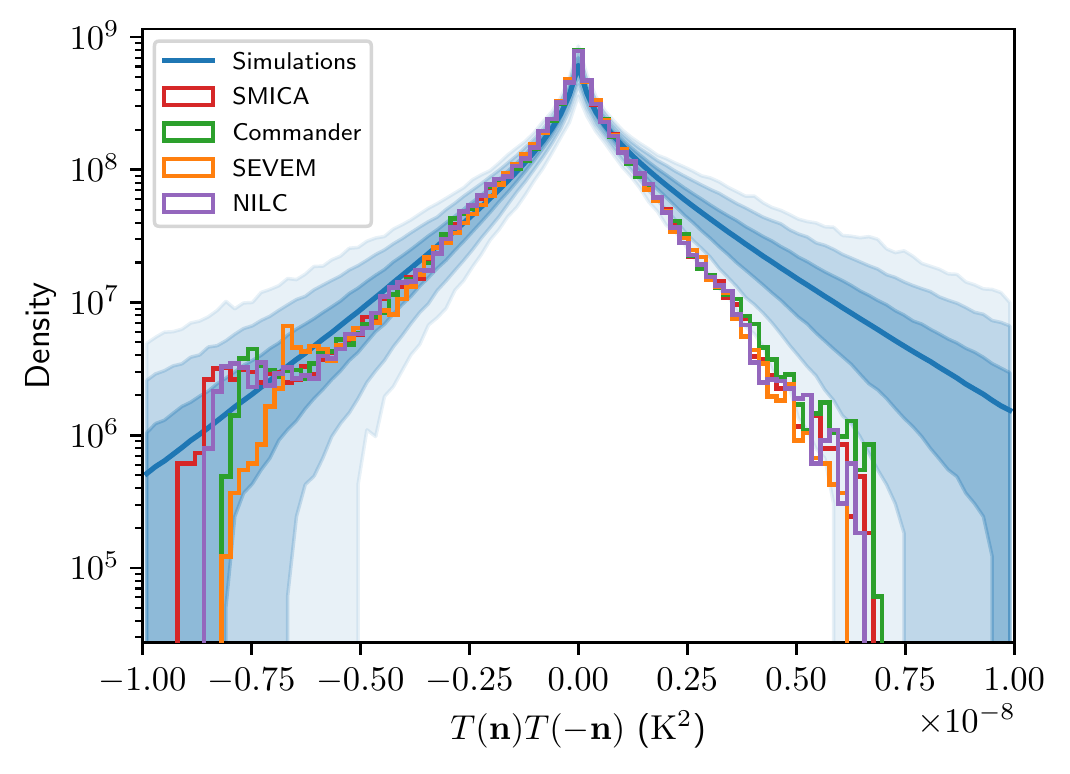}}
    \centerline{    
    \includegraphics[width=0.4\textwidth]{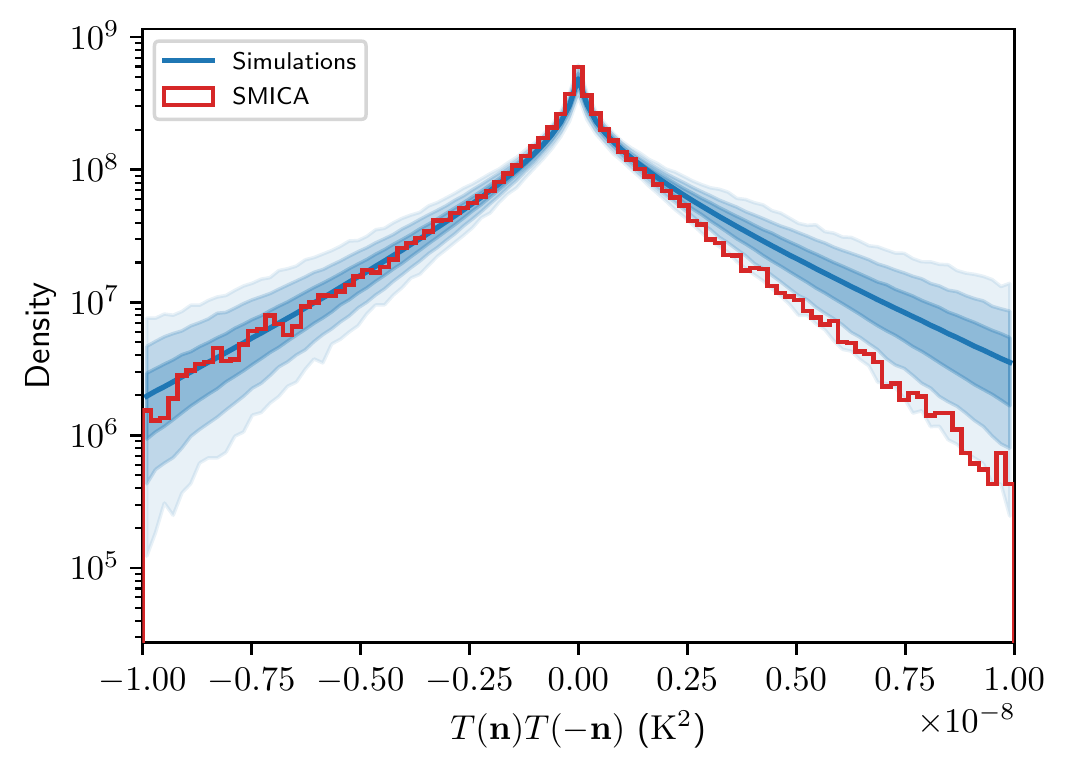}}   
    \caption{Top: the distribution function for number of counts(density)
    versus amplitude of $Z({\bf n})$- map with $\Theta=5^\circ$. The colored
    curves correspond to SMICA, Commander, SEVEM and NILC maps from Planck 2018 data release.\\
    Bottom: the same as top, but for $\Theta=2.5^\circ$. The blue solid line
    indicate the average distribution of 1000 Gaussian simulations. The light blue zones correspond to 68$\%$, $95\%$ and $99.7\%$ of realisations.}
    \label{fig2}
\end{figure}

In the pixel domain, the symmetric ($S({\bf n})$) and asymmetric ($A({\bf n})$) parts of the temperature anisotropy map are:
\begin{equation}
    S({\bf n}) = \frac{T({\bf n}) + T(-{\bf n})}{2};\quad
    A({\bf n}) = \frac{T({\bf n}) - T(-{\bf n})}{2}.
\label{eq:eq1}
\end{equation}
where ${\bf n}$ is a unit vector pointing to each pixel of the map. 
From equation~(\ref{eq:eq1}) one can define the following estimator of symmetry or
asymmetry of the temperature map \citep{Creswell:2021eqi}:
\begin{equation}
    Z({\bf n}) = T({\bf n}) T(-{\bf n})=S^2({\bf n})-A^2({\bf n}).
    \label{eq:eq2}
\end{equation}
Thus, for each pixel of the temperature map, a positive amplitude of the function $Z({\bf n})$ means dominance of symmetric component, while negative $Z({\bf n})$ corresponds to dominance of asymmetric component.
In figure~\ref{fig1} we plot $Z({\bf n})$ maps derived from Planck 2018 SMICA map
with Gaussian smoothing $\text{FWHM} = \Theta=2.5^\circ$ and $\Theta=5^\circ$.
In this figure the black solid line indicates the ring of attraction---the direction on the sphere orthogonal to the direction of the kinematic dipole (the axis of evil).
An important feature of figure~\ref{fig1} (top panel)  
is the presence of two pairs of very strong high negative peaks of $Z({\bf n})$ (labelled 1a/1b and 2a/2b) and about twenty negative peaks with smaller amplitudes, mainly localized within the area $|b|\le 30^\circ$ in Galactic coordinates. These peaks with $Z({\bf n})<0$ belong to asymmetric component of the signal and 
have the following Galactic coordinates  $(l, b)$:
\begin{align*}
    &\mathrm{1a}: (212^\circ, -21^\circ), \quad &&\mathrm{1b}: (32^\circ, 21^\circ)\\
    &\mathrm{2a}: (332^\circ, -8^\circ), \quad &&\mathrm{2b}: (152^\circ, 8^\circ)
\end{align*}
As it is seen from $Z({\bf n})$ with $\Theta=5^\circ$, the pair 2a/2b has
the Galactic longitude $|b|=8^\circ$. If the size of any Galactic masks exceeds this threshold, this region will make no contribution to the resulting the parity asymmetry.
A very important feature of the $Z({\bf n})$ map with $\Theta=5^\circ$ and $\Theta=2.5^\circ$ is related to the structure of the negative peaks shown in figure~\ref{fig2}.

The high amplitude peaks 1a/1b at low resolutions $\Theta=5^\circ$ reveal inner peak-like structure when we go to high resolution $\Theta=2.5^\circ$. In addition to amplification of the amplitudes of negative and positive peaks well above and below the ring of attraction, we can see inner structure of the 1a/1b and 2a/2b zones in the form of clustered peaks.

The pixelized $Z({\bf n})$ map can be converted into a distribution function of the number of counts versus amplitude $Z({\bf n})$.
We show this distribution in figure~\ref{fig2} for $\Theta=5^\circ$ (top) and $\Theta=2.5^\circ$ (bottom). These distributions reveal the following tendency. For low resolution $\Theta=5^\circ$, the $Z({\bf n})$ map has a bump in the distribution at $Z({\bf n}) \simeq -0.75 \times 10^{-8} \mathrm{K}^2$, while for positive
$Z({\bf n})\simeq 0.75 \times 10^{-8} \mathrm{K}^2$ we see an absence of the counts at the level of $3\sigma$. The bump at $Z({\bf n})\simeq -0.75 \times 10^{-8} \mathrm{K}^2$ and $\Theta=5^\circ$ is a common feature for the SMICA, Commander and
NILC maps. For SEVEM its position is slightly shifted and  corresponds to $Z({\bf n})\simeq -0.6 \times 10^{-8} \mathrm{K}^2$. For all the maps, the origin of the bump is associated with the peaks 1a/1b and 2a/2b.

For $\Theta=2.5^\circ$ the bump of the distribution function disappears, and the data follows the shape of Gaussian simulations. At the same time, for
both $\Theta=2.5^\circ$ and $\Theta=5^\circ$, we have a deficit of symmetric peaks responsible for the parity asymmetry. However, as we have seen from figure~\ref{fig1} at $\Theta=2.5^\circ$ there is decay  of  peaks 1a/1b and 2a/2b at $\Theta=5^\circ$ (adjusted to RA) in to the clusters of peaks with smaller scales.

\subsection{Theoretical shape of distribution}

Suppose $T({\bf n})$ is a realisation of a statistically isotropic Gaussian field.
Then, $T(-{\bf n})$ is also Gaussian-distributed.
However, these two quantities  are not independent random variables due to correlations in the pixel domain.
The precise details of this correlation are determined by the power spectrum and the smoothing angle $\Theta$.

For our purposes, it is sufficient to consider the Pearson cross-correlation coefficient of $T({\bf n})$ and $T(-{\bf n})$:
\begin{eqnarray}
\label{eq:rho}
\rho = \mathrm{Corr}(T({\bf n}), T({-\bf n}) = \frac{\displaystyle \int T({\bf n}) T({-\bf n}) \, d{\bf n}  }{\displaystyle \int T({\bf n})^2 \, d{\bf n}} .
\end{eqnarray}
We have taken that the mean is subtracted, $\int T({\bf n}) \, d{\bf n} = 0$. In realty the integrals are calculated as sums over all available pixels.
For given $\rho$, the distribution function of $Z({\bf n})$ has a form \cite{NADARAJAH2016201,gaunt}:
\begin{equation}
    \label{eq:P(Z)}
    \mathcal{P}(Z') = \frac{1}{\pi \sqrt{1 - \rho^2}} \exp(\frac{\rho Z'}{1 - \rho^2}) K_0 \qty(\frac{|Z'|}{1 - \rho^2}), 
\end{equation}
where $K_0$ is the 0-th order modified Bessel function of the second kind
and $Z' = Z / \mathrm{var}(T)$ is the rescaled $Z$ to unit variance.

The distribution function $\mathcal{P}(Z')$ is asymmetric for positive and
negative $Z$. In order to estimate the coefficient of correlation $\rho$ for the best fit Planck 2018 $\Lambda$CDM cosmological model and smoothing angle $\Theta=5^\circ$ we run 1000 simulations for statistically isotropic random Gaussian realizations and get the mean value and $1\sigma$ standard deviation:
$\rho=0.105\pm 0.15$. The actual value of $\rho$ for the distribution function
presented in figure~\ref{fig2} varies from $\rho=-0.136$ for pixels with $\mathcal{P}(Z')>5 \times 10^7$ and $Z<0$ down to $\rho\simeq -0.35$ at $\mathcal{P}(Z')\simeq 3 \times 10^6$, in the domain of the bump. 

\subsection{Asymmetry estimator}

\begin{figure*}[t]
    \centering
    \includegraphics[width=0.45\textwidth]{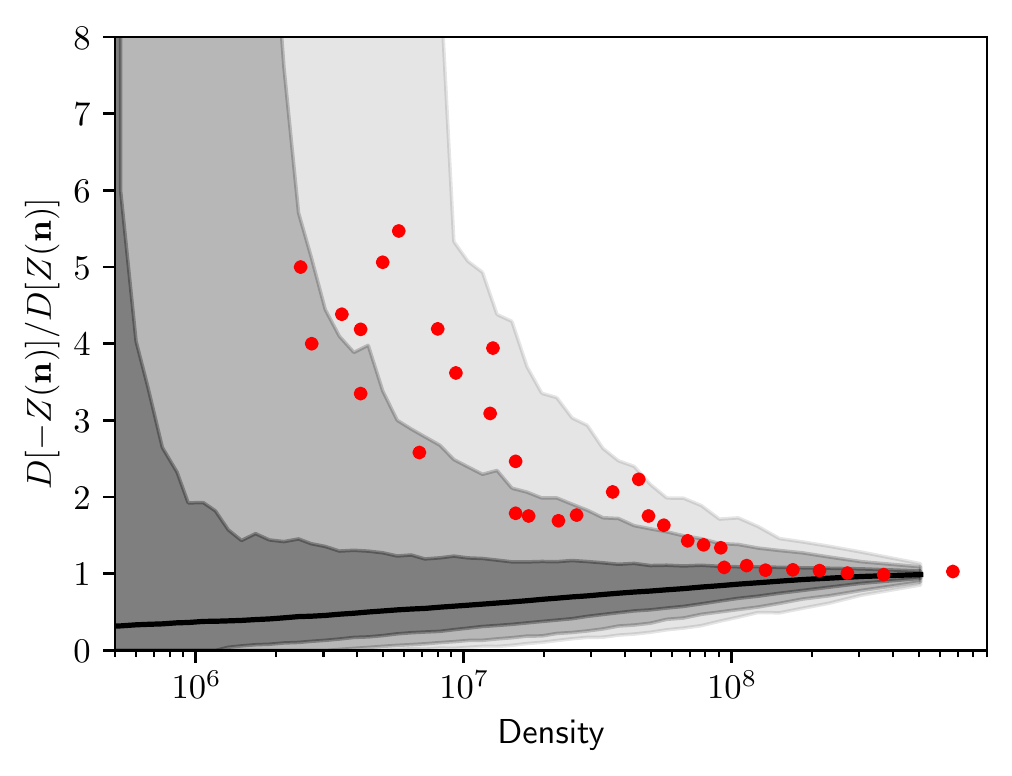}
    \includegraphics[width=0.45\textwidth]{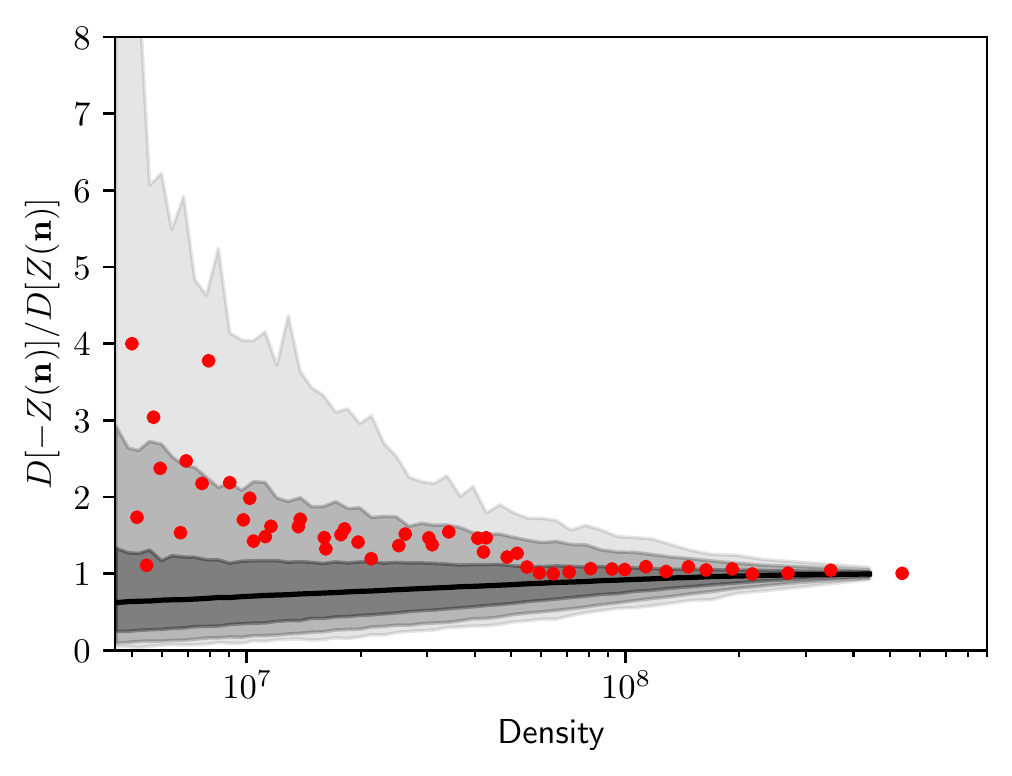} 
    \caption{Ratio of the density on the left to the density on the right, as a function of the position of the density, for smoothing angle $5^\circ$ (left) and $2.5^\circ$ (right).
    This sort of estimator is therefore a measure of the asymmetry of the distribution function, at each amplitude. 
    The black solid line indicates the average of 1000 Gaussian simulations, and the grey zones correspond to 68$\%$, $95\%$ and $99.7\%$ of realizations.
    }
    \label{fig3}
\end{figure*}

In order to estimate the significance of asymmetry we use the following estimator, based on the properties of the distribution function $\mathcal{P}(Z')$. The range of the distribution function is divided into 50 bins, and in each bin is determined the corresponding value of $Z$ for asymmetric $D(Z_a)$ and symmetric $D(Z_s)$ amplitudes. We then introduce the   ratio 
\begin{equation}
R(\mathcal{P}(Z))=\frac{D(Z_a)}{D(Z_s)}
\label{ratio}
\end{equation}
as a measure of asymmetry between asymmetric and symmetric tails of distribution function. The same estimator is applied to realizations of a random Gaussian signal. The results of comparison  are shown in figure~\ref{fig3}, where lines of different shades of gray correspond to 68$\%$, 95$\%$ and 99.7$\%$  of realizations. We present in figure~\ref{fig3}  two variants of the distributions for $\Theta=5^\circ$ and $\Theta=2.5^\circ$ At resolution $\Theta=5^\circ$ in the bump area, $R$ departs at the $3\sigma$ level. For resolution $\Theta=2.5^\circ$ , the anomaly level is reduced to around $2\sigma$.

\begin{figure}[t]
    \centering
    \includegraphics[width=0.45\textwidth]{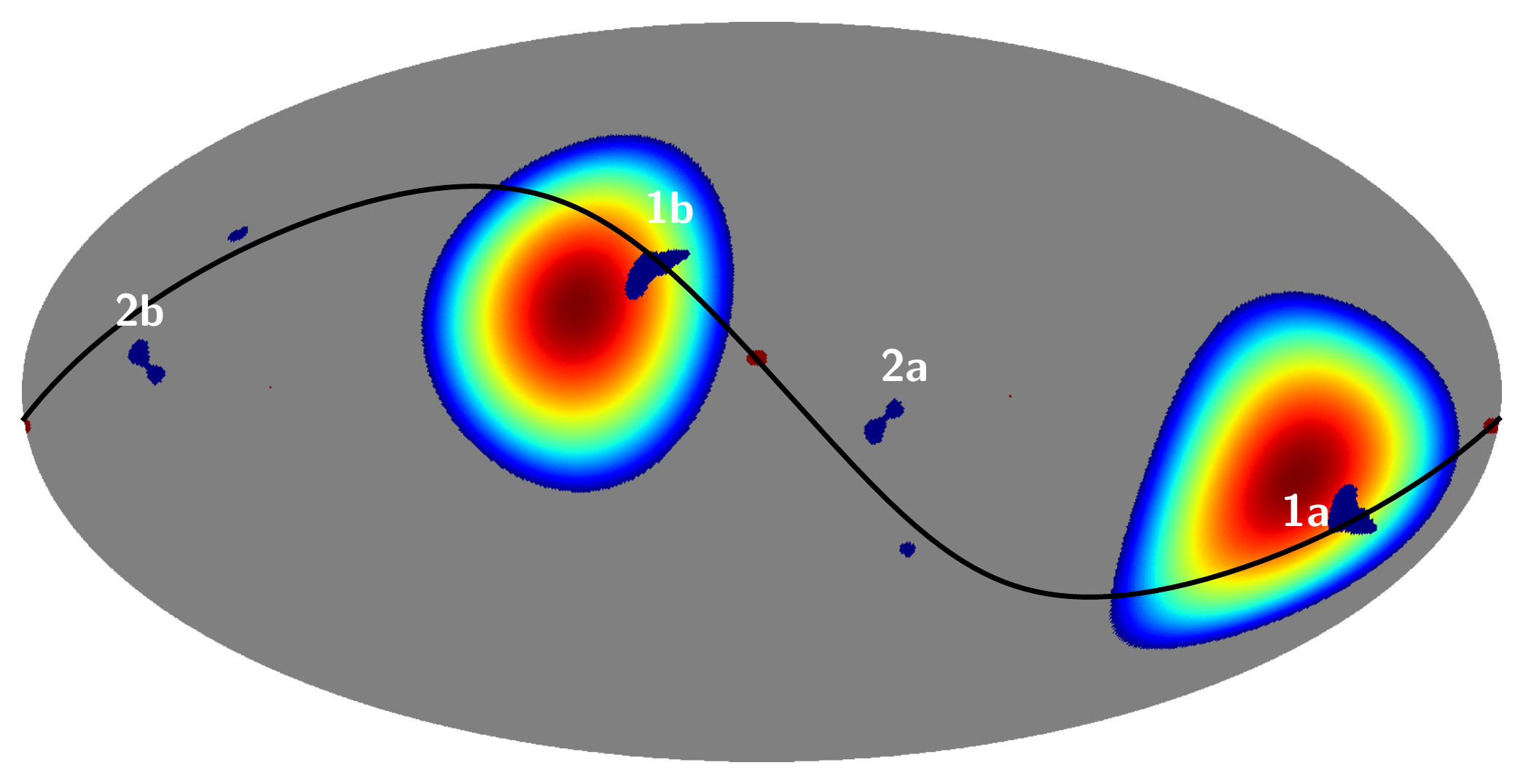}
    \includegraphics[width=0.45\textwidth]{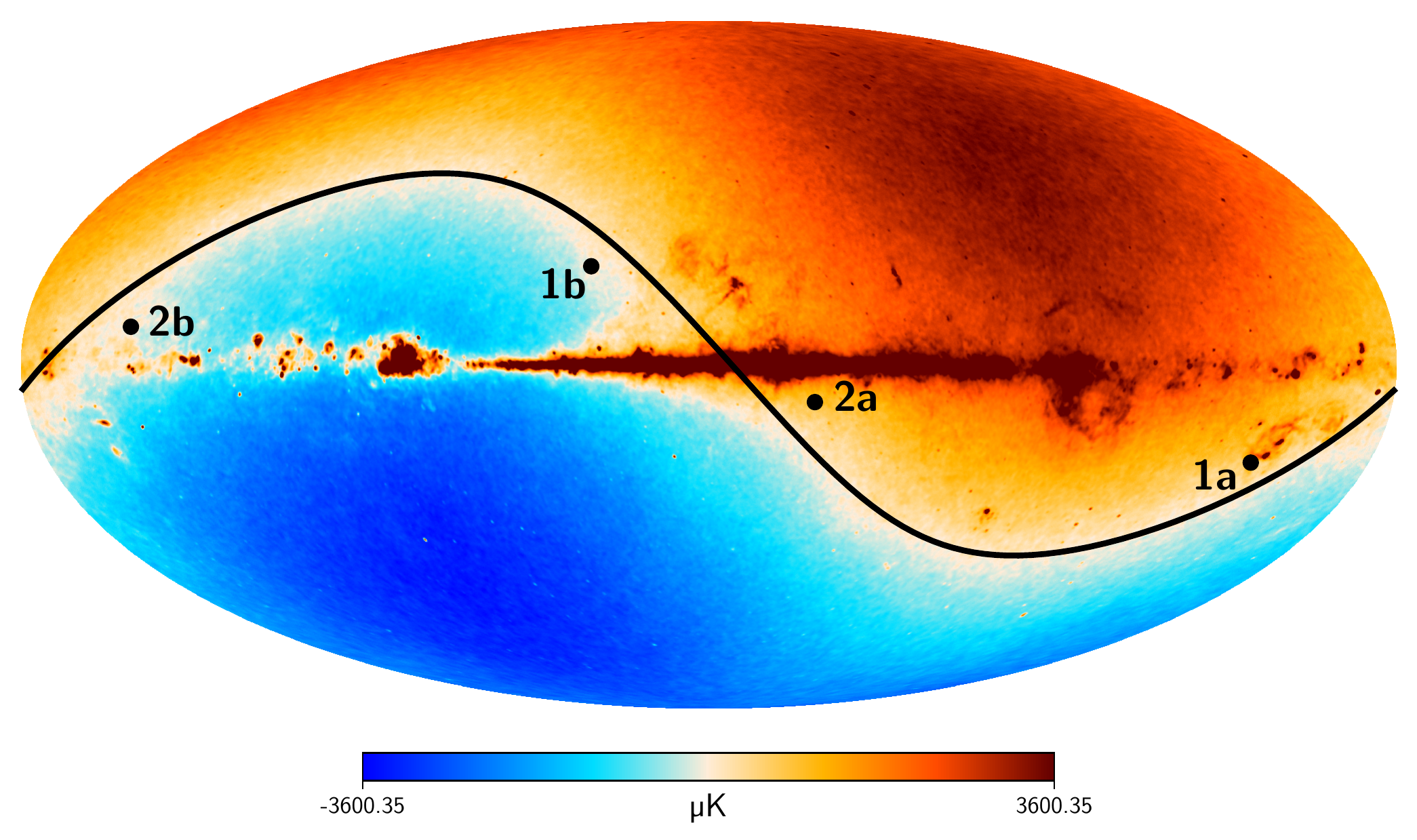}   
    \caption{Top panel: a posterior distribution of the direction of the dipole modulation (the colored countors) in combination with RA and
    1a/1b and 2a/2b peaks responsible for maximum of the parity asymmetry.
    The right contours correspond to the maximum of CMB power, the left contours indicate the minimum. Bottom panel: the 30 GHz Planck map with
    kinematic dipole unsubtracted (source: BeyondPlanck 2020 data release \cite{Andersen:2020jfq}).}
    \label{fig4}
\end{figure}

The smoothing angle $\Theta$ is related to the characteristic multipoles through $\Theta\simeq{100^\circ}/{\ell}$. For $\Theta=5^\circ$ the corresponding 
multipoles are localised around $\ell\sim 20$ to $30$, while for $\Theta=2.5^\circ$
we have $\ell\sim 40$ to $50$. Actually, the decrease of significance for $\ell>30$
is in agreement with results in \citep{Kim_2012,IS2015,IS2018,Gruppuso_2010,Gruppuso:2017nap}.

\section{Common directions of dipole modulation and parity asymmetry}

In this section we investigate the problem of possible common origin of
parity asymmetry, the dipole modulation of the CMB and properties of kinematic dipole as a generator of anomalies. 
We will start with the model of dipole modulation, proposed in \citep{Eriksen_2004b} for
explanation of the CMB power asymmetry. 

According to \citep{Hansen_2004,Eriksen_2007,Hoftuft_2009} the observable temperature anisotropy $T_{obs}({\bf n})$ related to a primordial statistically isotropic and Gaussian signal $T_g({\bf n})$ as:
\begin{equation}
T_{obs}(\mathbf{n})=\left(1+D(\mathbf{m}\vdot\mathbf{n})\right)T_g(\mathbf{n})
\label{dipole}
\end{equation}
where ${\bf m}$ is unit vector in the direction of dipole modulation, ${\bf m}\vdot{\bf n}$ denotes the dot product, and $D\simeq 0.07$ is the amplitude.  Following \cite{Hoftuft_2009}, in Galactic coordinates, the vector ${\bf m}$ points towards the direction $(l,b) = (224^\circ,-22^\circ)\pm 24^\circ$.

This model equation~(\ref{dipole}) assumes that only the primary Gaussian CMB signal is modulated, but not foregrounds or any non-cosmological signals. This means that the direction of the dipole ${\bf m}$ should not correlate with these sources of anisotropy. Moreover, if the dipole modulation and the parity asymmetry are statistically independent, their characteristic directions should not correlate either.

In reality, as it is seen from figure~\ref{fig4}, all these criteria do not appear to hold. The most significant 1a/1b peaks of $Z({\bf n})$, which correspond to
$3\sigma$ anomaly in figure~\ref{fig3}, coincide with the position of the zone of the most probable direction of the dipole ${\bf m}$, and the direction of ring of attraction crosses both the positions of these peaks and the zone of the most probable orientation of vector ${\bf m}$.

\subsection{Significance estimation}

\begin{figure}[t]
    \centering
    \includegraphics[width=0.48\textwidth]{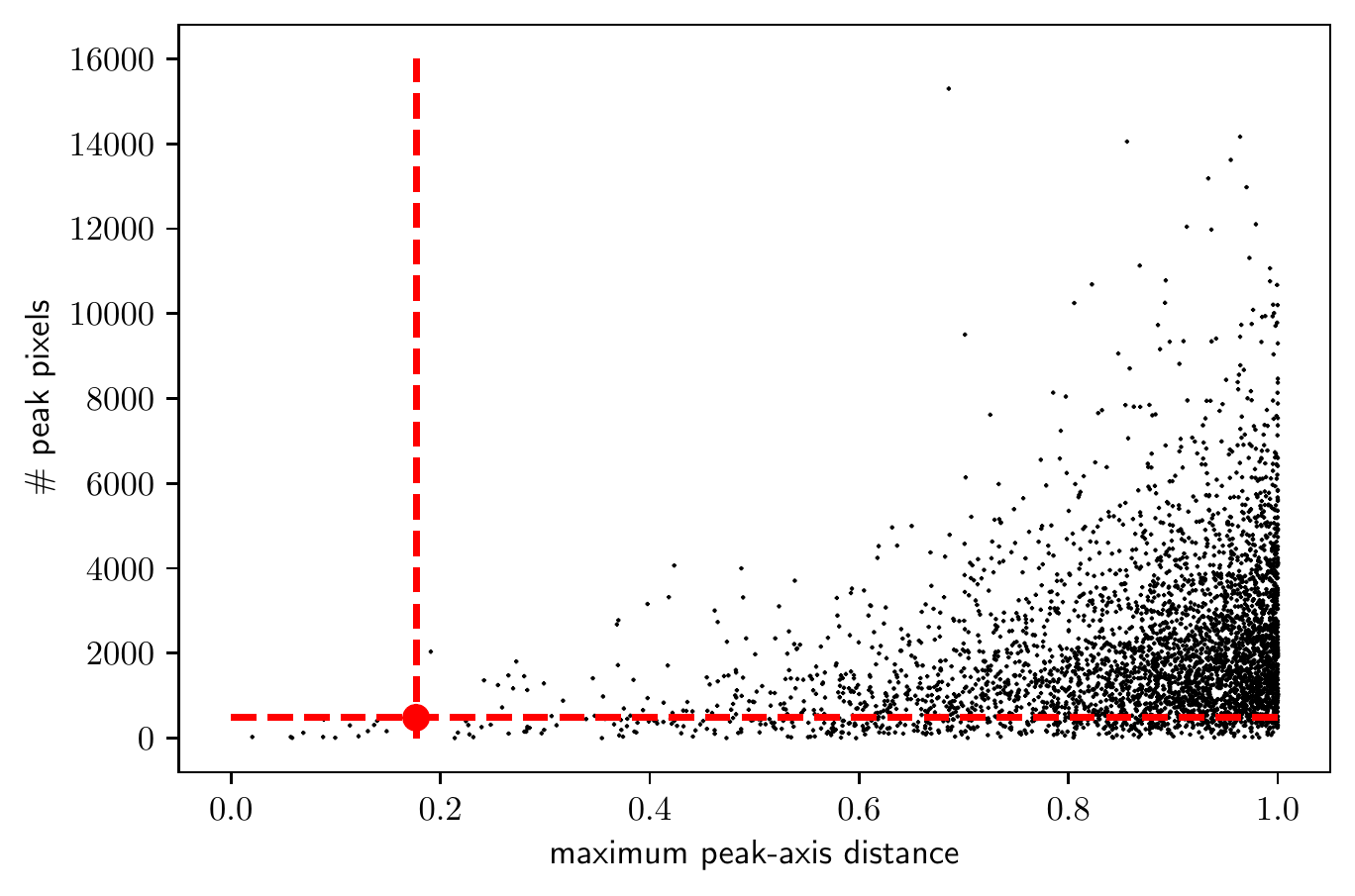}
    \caption{The values of $\max_{|Z(\mathbf{n})| \geq \nu} (\mathbf{m} \cdot \mathbf{n})$ and the number of peak pixels for 4000 simulations (black dots), compared to SMICA (red dot).}
    \label{fig:significance}
\end{figure}

To quantify this correlation, in this section we estimate the significance of the overlap of the peaks of the estimator of asymmetry (equation~(\ref{eq:eq2})) with the ring of attraction, which are supposed to be independent according to the null hypothesis.
As seen in figure~\ref{fig1}, there is strong coincidence of the peaks of the $Z(\mathbf{n})$ map with the RA, which is measured by the estimator
\begin{equation}
   d= \max_{|Z(\mathbf{n})| \geq \nu} (\mathbf{q} \cdot \mathbf{n}), 
\end{equation}
where $\mathbf{q}$ is the unit vector in the direction of the kinematic dipole, $\nu$ is the peak threshold, and the maximum is taken over all peak pixels $\mathbf{n}$ for which $|Z(\mathbf{n})| \geq \nu$.
Therefore this estimator gives us an upper bound on the distance between the peaks of $Z$ and the ring of attraction.

We also consider the complementary estimator which is the number of pixels where $|Z(\mathbf{n})| \geq \nu$, or equivalently the sky area of the peaks.

\begin{figure*}[t]
    \centering
    \includegraphics{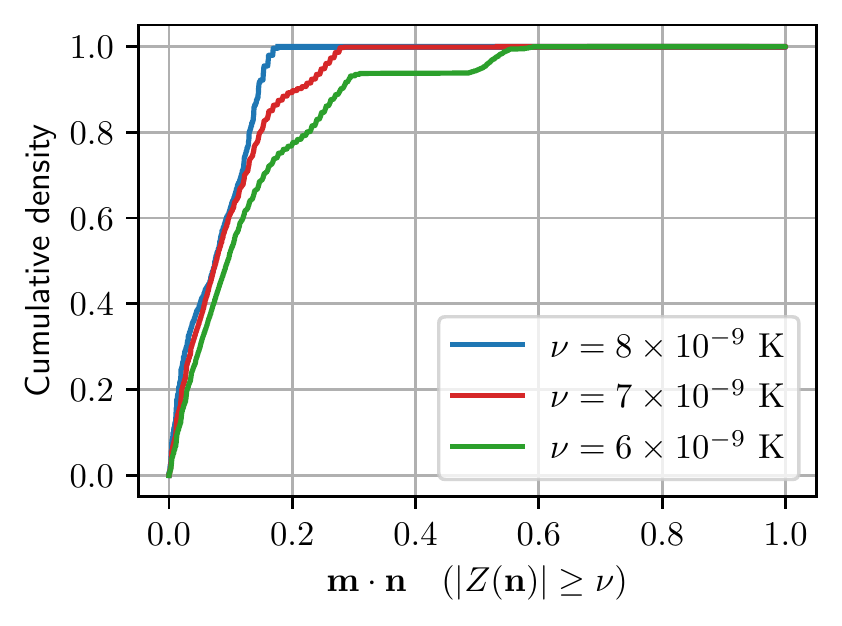}
    \includegraphics{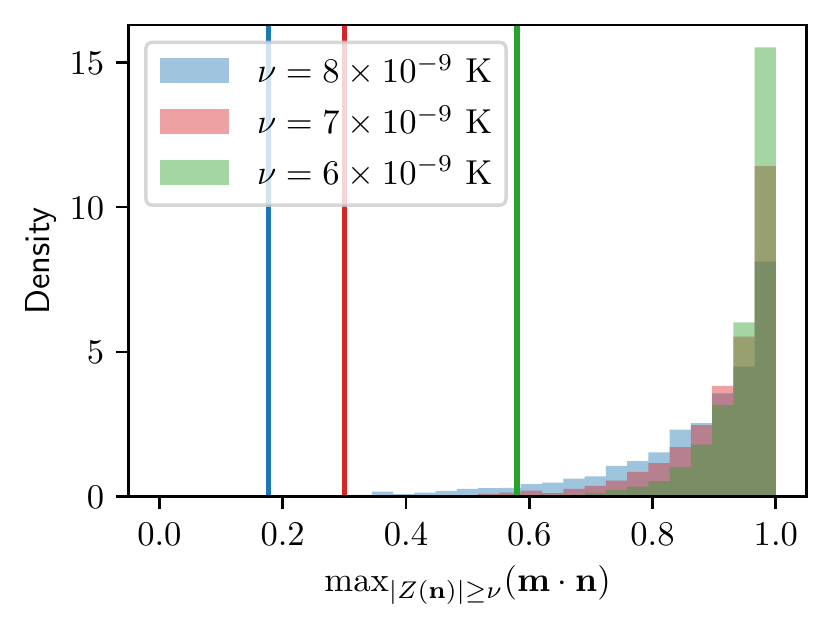}
    \caption{Left panel: CDFs of the peak--ring distance, for three different choices of peak threshold. The concentration of $Z$ peaks near the ring is remarkably stable. Right panel: the corresponding PDFs, and the values measured with SMICA shown in vertical bars.}
    \label{fig:cdf}
\end{figure*}

 \begin{figure*}[!t]
    \centerline{
    \includegraphics{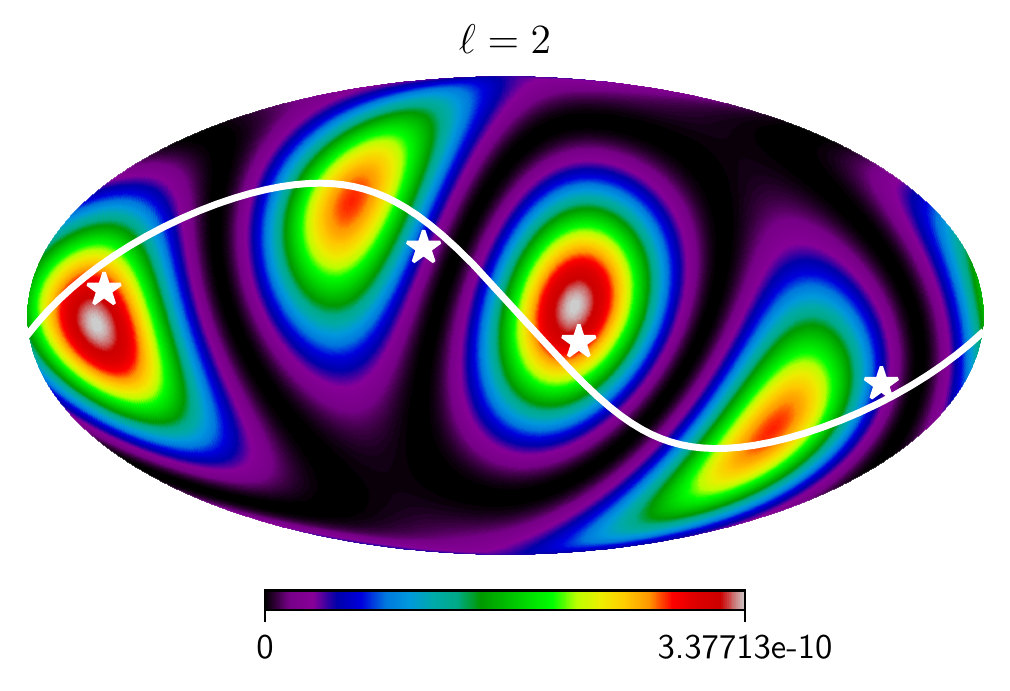}
    \includegraphics{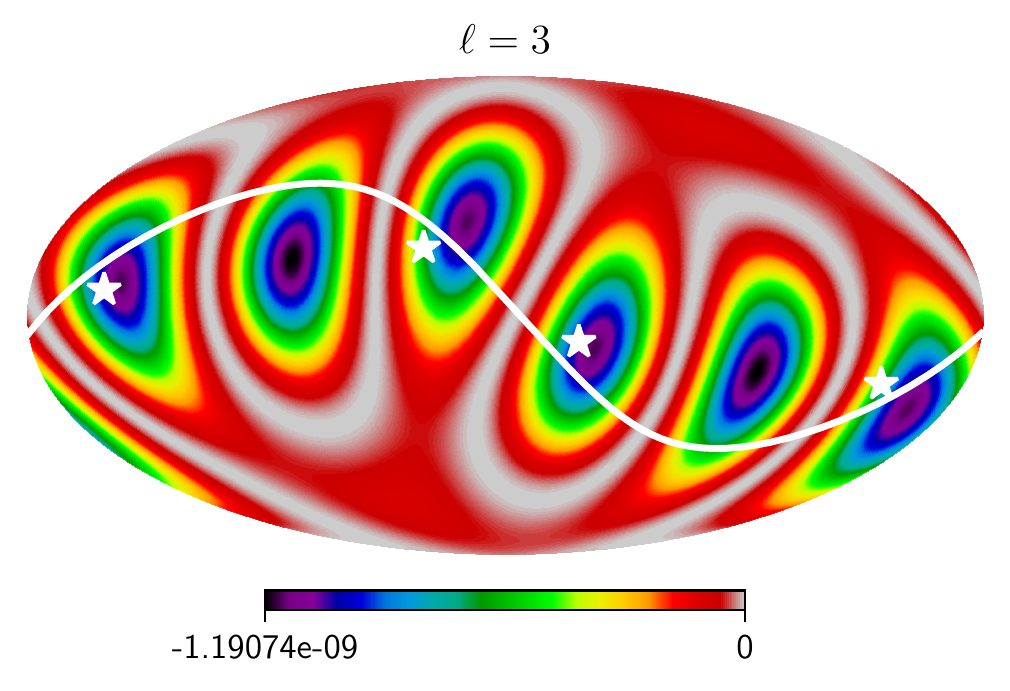}}
  \centerline{
    \includegraphics{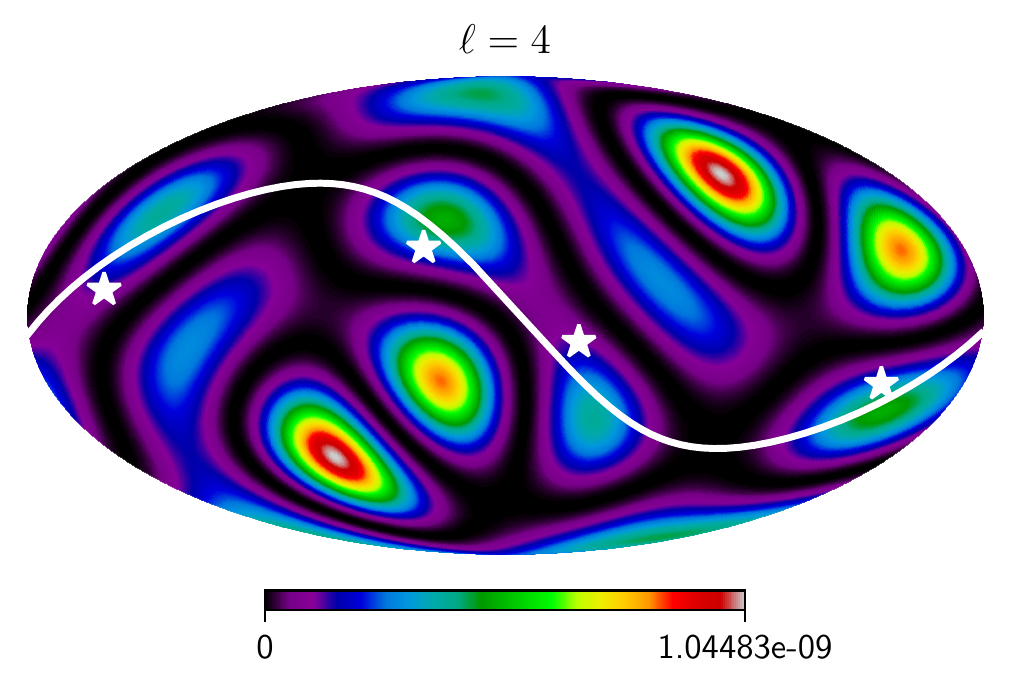}
    \includegraphics{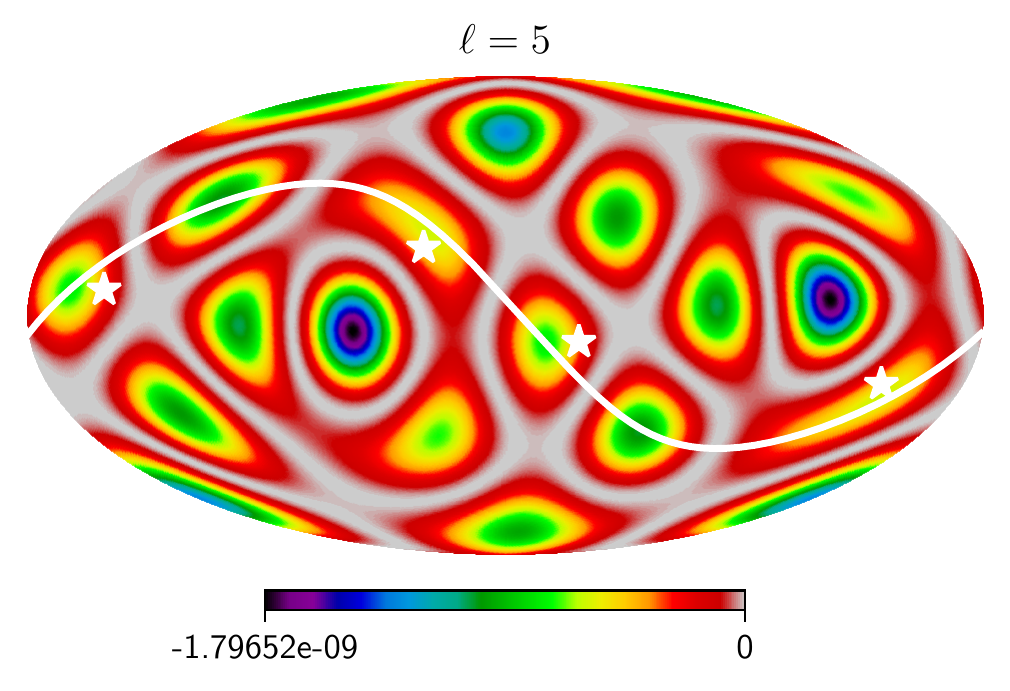}}
      \centerline{
    \includegraphics{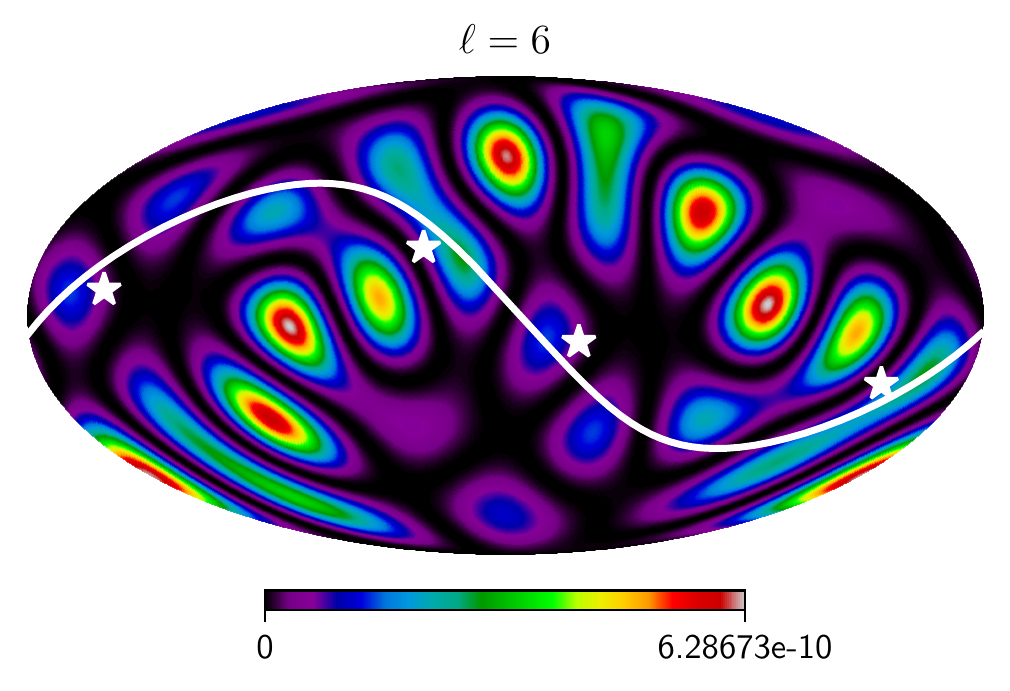}
    \includegraphics{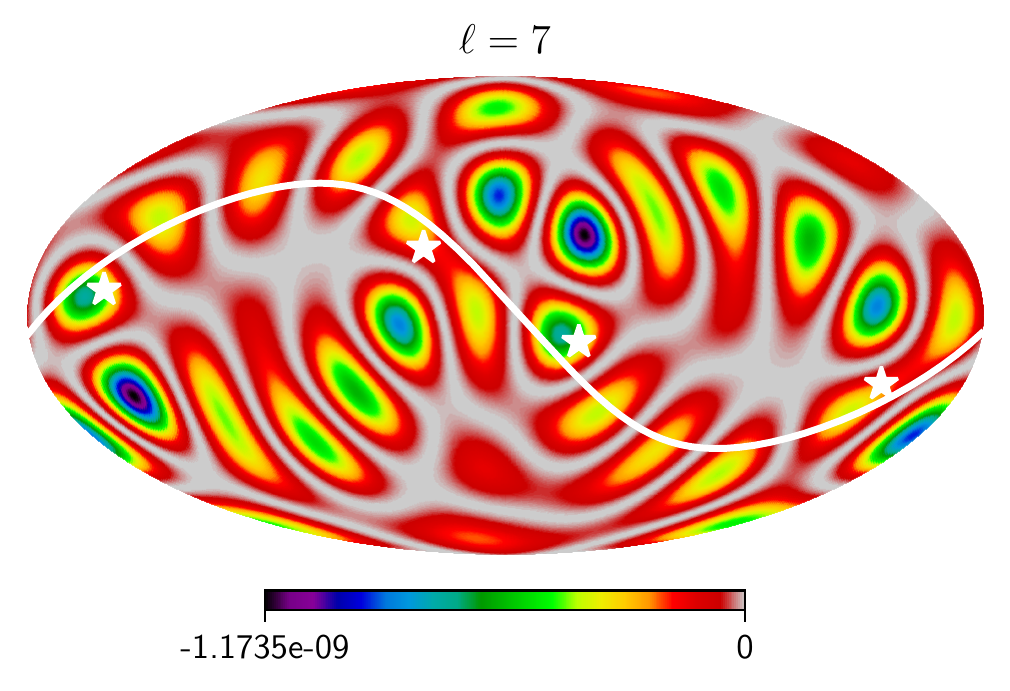}}
    \caption{The left column: $Z_{\ell}(\mathbf{n})$-maps for even ($\ell=2,4,6$) multipoles. The 
    right column: $Z_{\ell}(\mathbf{n})$-maps for odd ($\ell=3,5,7$) multipoles. The white solid line indicate RA. The stars marks the position of the peaks 1a/1b and 2a/2b.}
    \label{fig7}
\end{figure*}

In figure~\ref{fig:significance}, we show, for 4000 Gaussian simulations, the values of these two estimators on the $x$ and $y$ axes.
Needless to mention, these two estimators are not independent.
The simulations occupy a triangular region in the phase space, which is not surprising, because if a particular realization has fewer peaks, then it is allowed greater random fluctuations in the location of those peaks.
This correlation must be considered when estimating the significance of SMICA, which is shown by the red point.
Among all 4000 simulations, there are approximately 20 with a smaller maximum peak--ring distance.
However, most of these have only a very small number of peak pixels.
Among Gaussian simulations with equally many or more peak pixels than SMICA, there is found 1 in 2000 realizations with equal or lesser value of the $d$ estimator.

\subsection{Variation of threshold $\boldsymbol \nu$}

Figure~\ref{fig:significance} uses a peak threshold of $\nu = 8 \times 10^{-9}$ K.
This corresponds to a strict definition of the peak region---0.25\% of the sky for SMICA exceeds $|Z(\mathbf{n})| \geq 8 \times 10^{-9}$ (but on this point, note the general deficit of positive peaks for SMICA, visible in figure~2).
We also test different peak thresholds, reducing $\nu$ to $7 \times 10^{-9}$ K (corresponding to $f_\mathrm{sky} = 0.65\%$ and $6 \times 10^{-9}$ K (corresponding to $f_\mathrm{sky} = 1.15\%$).
In the left panel of figure~\ref{fig:cdf}, the cumulative density function (CDF) of $\mathbf{m} \cdot \mathbf{n}$ for pixels $\mathbf{n}$ obeying $|Z(\mathbf{n})| \geq \nu$ for SMICA is shown for each of these thresholds.

Note that all three thresholds result in visible strong concentration of peaks near to the ring of attraction.
When the threshold is increased, the attraction increases.
In the right panel of figure~\ref{fig:cdf} is shown the distribution functions of the estimator $d$ expected for isotropic Gaussian simulations with the three choices of threshold.
In all cases we have a very significant departure of SMICA.

\begin{figure}[t]
    \centering
    \includegraphics[width=0.4\textwidth]{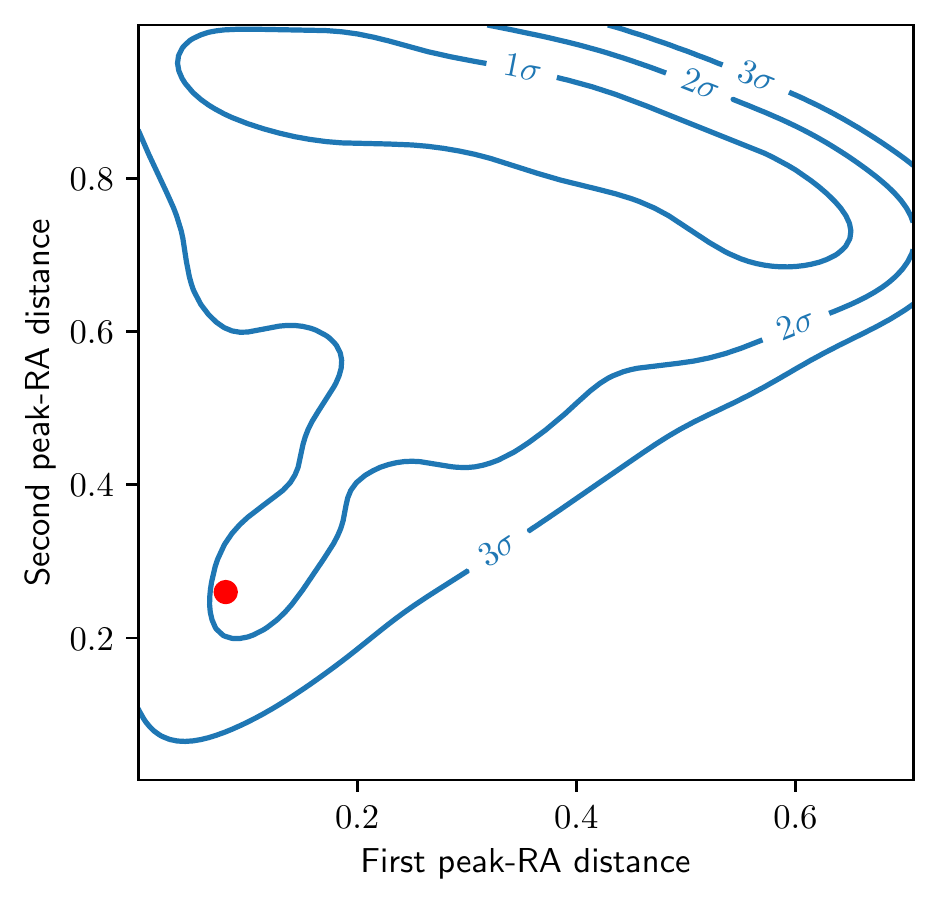}
    \caption{Distance between the two pairs of peaks of the quadrupole to the RA. SMICA is shown by the red dot.}
    \label{fig8}
\end{figure}

\section{Low multipoles and RA}
 
The problem of the statistical properties of the low multipoles (quadrupole and octupole) has a long history. It  started with an abnormally low quadrupole amplitude in the COBE data \citep{Bennett:1996ce}, then
a low quadrupole was recorded in the data from WMAP and Planck \citep{Spergel:2003cb,10.1046/j.1365-8711.2003.06940.x}. Besides,
the literature intensively discusses the anomalous mutual orientation of the quadrupole and octupole and the existence of the ``axis of evil''---the common direction for
these two components in the multipole vector approach \citep{Copi_2004,Schwarz_2004}.
In this section we return to the discussion of the quadrupole-octupole problem
within the $Z(\mathbf{n})$ approach.
The main question is how these components are related to the RA direction.
 
 \begin{figure*}[t]
    \centering
    \includegraphics[width=0.32\textwidth]{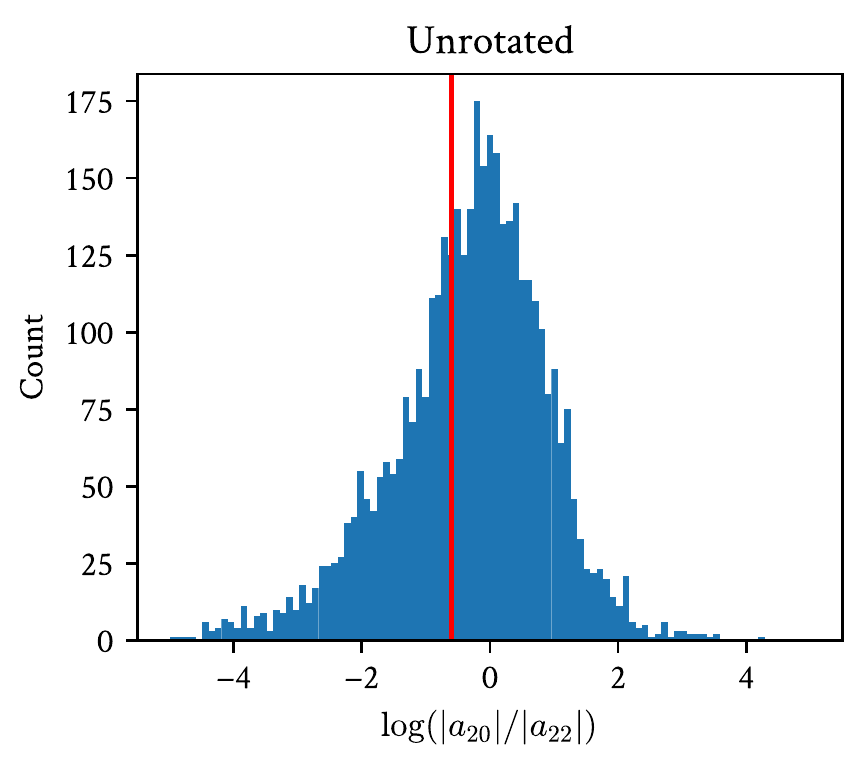}
    \includegraphics[width=0.32\textwidth]{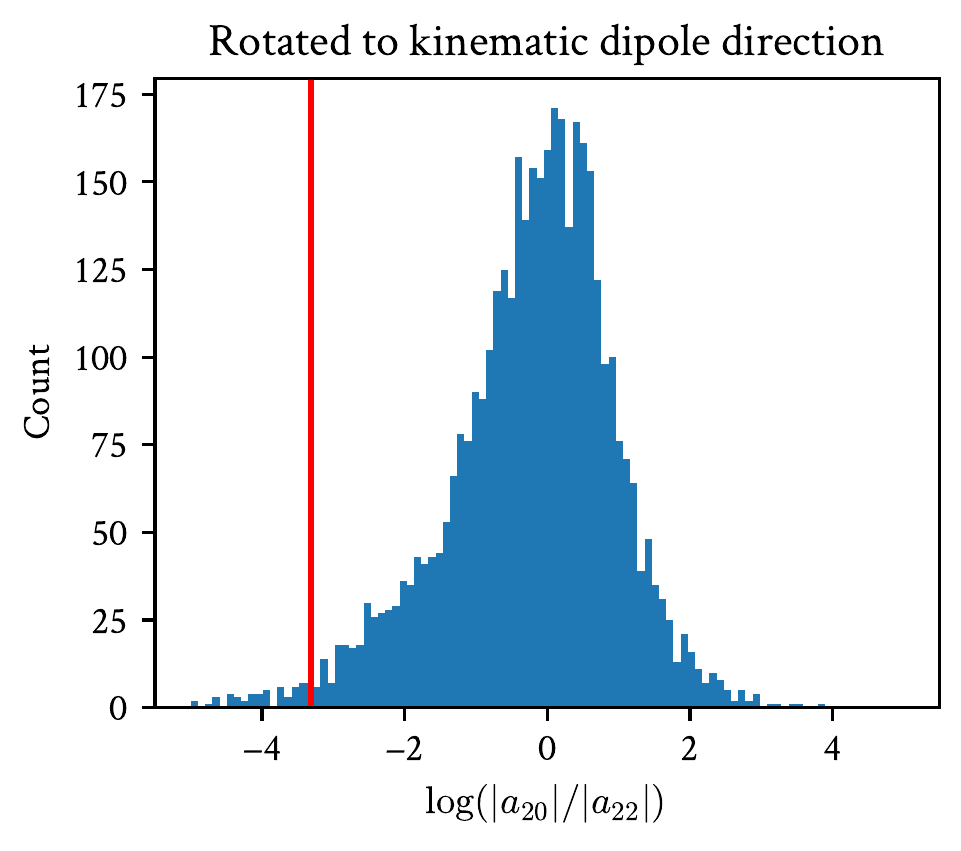}
     \includegraphics[width=0.32\textwidth]{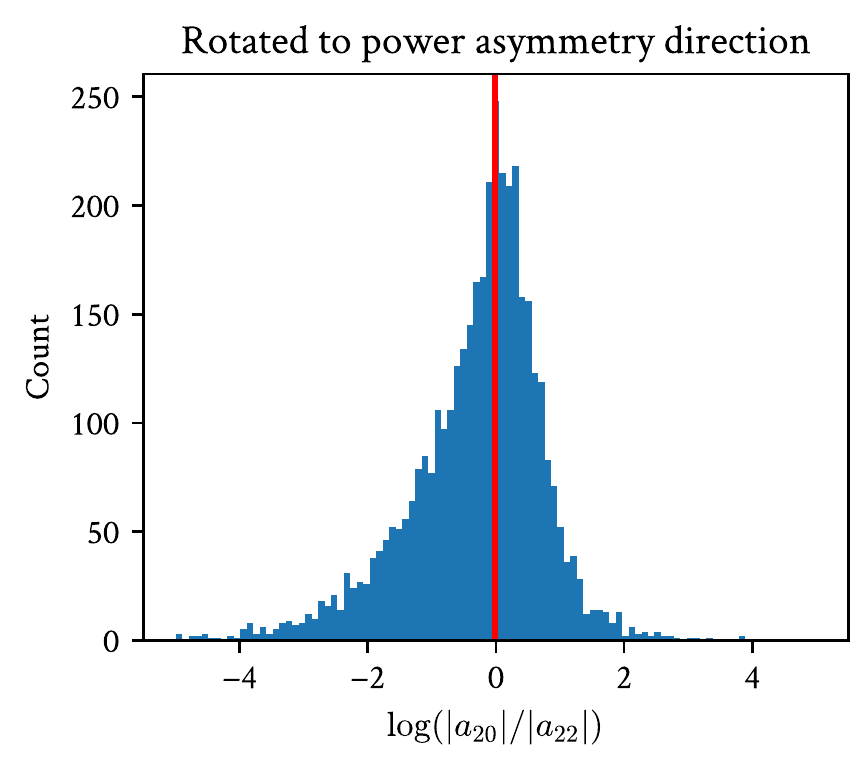}
    \caption{The amplitude of $\Gamma$ for Galactic coordinates without rotation (left), after rotation in the direction of the kinematic dipole (middle) and for the direction of the dipole modulation (right). For the comparison we show histograms for random Gaussian simulations.}
    \label{fig9}
\end{figure*}

As previously, we will use the SMICA temperature map and decompose it to particular maps for each multipole $\ell=2,3, \dots, 7$. Then we convert these maps into symmetric (even $\ell$) and asymmetric (odd $\ell$)  $Z$-maps and include in these maps the position of RA. Our natural expectation is that
RA will attract the $Z$-peaks of the odd modes and should not be influenced on the even modes. In reality the tendency is much more complicated, as it is seen from figure~\ref{fig7}. The pure RA effect is visible only for the octupole for all the peaks of the corresponding map $Z_3$. For the quadrupole the RA direction is still presented as a common factor for all the peaks. The highest two peaks coincide with peaks 2a/2b of the map $ Z (\mathbf{n}) $ from figure~\ref{fig1}, while the peaks 1a/1b do not coincide even with subdominant peaks of the quadrupole.
The significance of this effect can be estimated simply based on the distance between the two pairs of peaks, and RA. In figure~\ref{fig8} is shown the contours in the 2-dimensional space of the distances between the pairs of peaks of the quadrupole, based on random Gaussian simulations. In this way we see that the quadrupole peak alignment is significant at around the $2\sigma$ level.

It is important to note that, as visible in figure~\ref{fig7}, the attraction of peaks to the RA for multipoles $\ell \geq 4$ is not a property of the strongest peaks.
Instead, it is seen a series of subdominant peaks aligned with the RA.
In the case of $\ell = 6$ and $\ell = 7$, there are alignments of stronger peaks in another direction.
We would also like to point out the presence of strong peaks in the Galactic plane region, for example in $\ell = 5$.

The same level of quadrupole peculiarity can be obtained by rotating the coordinate system in given directions in the sky:
\begin{equation}
a_{\ell,0}(\mathbf{j})=\sum_{m'=-2}^{2}a_{\ell=2,m'}W^{\ell=2}_{m=0,m'}(0,\beta,\gamma)
\label{estimator4}
\end{equation}   
where $\mathbf{j} \equiv (\beta,\gamma)$ are the Euler angles, $a_{\ell,m}(\mathbf{j})$ are the coefficients after rotation of the reference system in the direction $\mathbf{j}$ , and
$W^{\ell}_{m,m'}$ is the Wigner rotation matrix.    

We will be interested in
two special cases---the direction of the kinematic dipole and the direction of the dipole modulation. In both cases, we will use the ratio $\Gamma=\frac{|a_{2,0}|}{|a_{2,2}|}$ as an estimator of abnormality and summarised the results in figure~\ref{fig9}.

The corresponding $p$-values for each particular direction are 0.43, 0.01, 0.45. This result is not surprising since the $p$-value critically depends on the estimator applied.
Thus, the estimator $\Gamma$ detects abnormality of the quadrupole at the same level, as in figure~\ref{fig8}. It is worth to note, that in terms of the $\Gamma$ estimator, the RA corresponds to $\Gamma\simeq 1$.

The most interesting information is coming from the analysis of the $Z$-maps of the odd multipoles.
For $\ell=5$ the highest peaks of $Z_5$ almost coincide with the Galactic plane and one may think that RA does not play an essential role for this mode. However, we want to pay attention to the sequence of subdominant peaks located strictly along the RA. The same feature is typical for the $\ell=7$ map in figure~\ref{fig7}. It could be thought that these matches are statistically insignificant and are a simple selection effect. However, we must not forget that in the CMB map it is these subdominant directions  after summation, that lead to the formation of high peaks 1a/1b and 2a/2b. This means that the presence of RA in the odd $\ell$ maps (which contribute to negative $Z(\mathbf{n})$) is a stable factor. For even $\ell$, except for the quadrupole and to a lesser extent $\ell = 4$ and $8$, the presence of RA is not observed so strongly in the maps.

\section{Discussion and conclusion}

In recent work \citep{Creswell:2021eqi}, it was argued that the CMB parity asymmetry can be investigated using a pixel-domain approach.
Under this investigation, in the present paper we found that the parity asymmetry is especially associated with anomalous density of antipodal peaks, associated to the directions (RA) orthogonal to the direction of the kinematic dipole.
Moreover, there may be a link between the parity asymmetry and the other low-$\ell$ anomaly, the dipole modulation asymmetry, whose direction can also be detected in the full-sky distribution of the parity asymmetry estimator.

To measure the significance of alignment to RA, we introduced the estimator which is, for a given threshold, the maximum distance from the RA of all pixels above that threshold.
Under the null hypothesis, only a few simulations  will have small maximum peak-RA distances, because the peaks will be randomly distributed on the sky.
For the actual SMICA data, this estimator is highly peculiar at about the 1 in 2000 level, reflecting the clustering of $Z(\mathbf{n})$ peaks near to the RA.
The same result can be achieved with different choices of the threshold.
We have used thresholds $6\times10^{-9}$ K, $7\times10^{-9}$ K, and $8\times10^{-9}$ K, which correspond respectively to sky areas of between 0.3\% and 1.15\% highest-$|Z|$ pixels on the sky.

However, the phenomenon of the RA is not simply restricted to the the peaks of the $Z(\mathbf{n})$.
As shown in figure 7, there is remarkable concentration of power from the $\ell = 2$ and $3$ (quadrupole and octupole) modes and subdominant correlated the low odd-$\ell$ modes aligned with the RA. 
This  alignment is robust against the choice of different masking strategies of the Galactic plane.
The results do not rule out the possibility that the parity asymmetry is partially associated with the Galactic plane, and peaks 2a/2b (with Galactic latitude $b = \pm 8^\circ$) could be removed in the masked analysis. 

Our last remark relates to the detected significance level of the anomalies, which ranges between 2 and 3$\sigma$. This is usually interpreted as an indication of the presence of anomalies that are statistically
mild compared to 5$\sigma$ threshold and may be an artifact of
Gaussian distribution. However, the results   of studies by \cite{Ben-David:2015fia,Ben-David:2015sia,Rahman:2021azv} of statistical
foreground anisotropies for both the Haslam synchrotron map and the  thermal dust outside the Galactic masks show that their statistical deviations from isotropy and Gaussianity are just within these limits of the confidence interval.

\begin{figure*}
    \centering
    \includegraphics[width=0.48\textwidth]{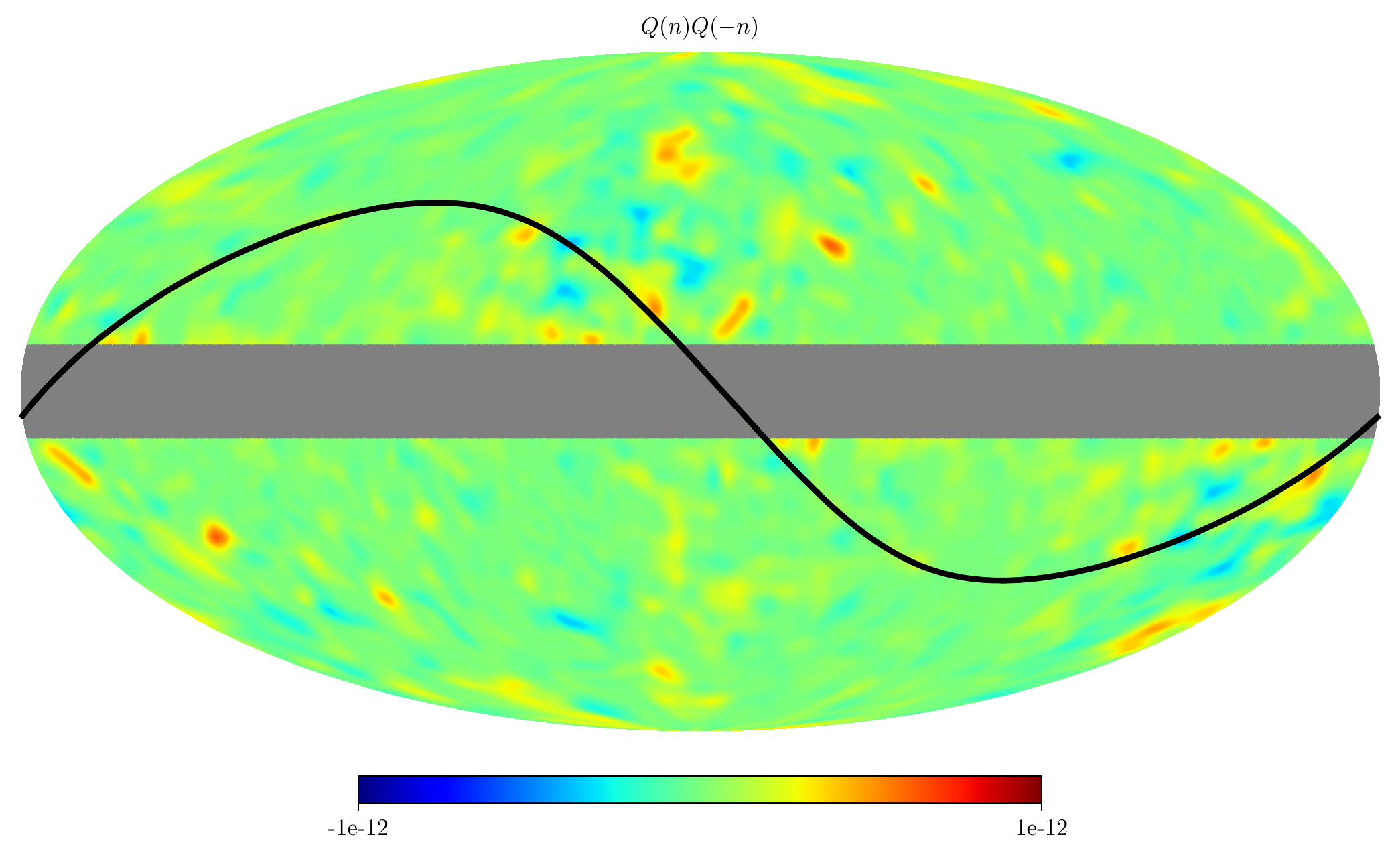}
    \includegraphics[width=0.48\textwidth]{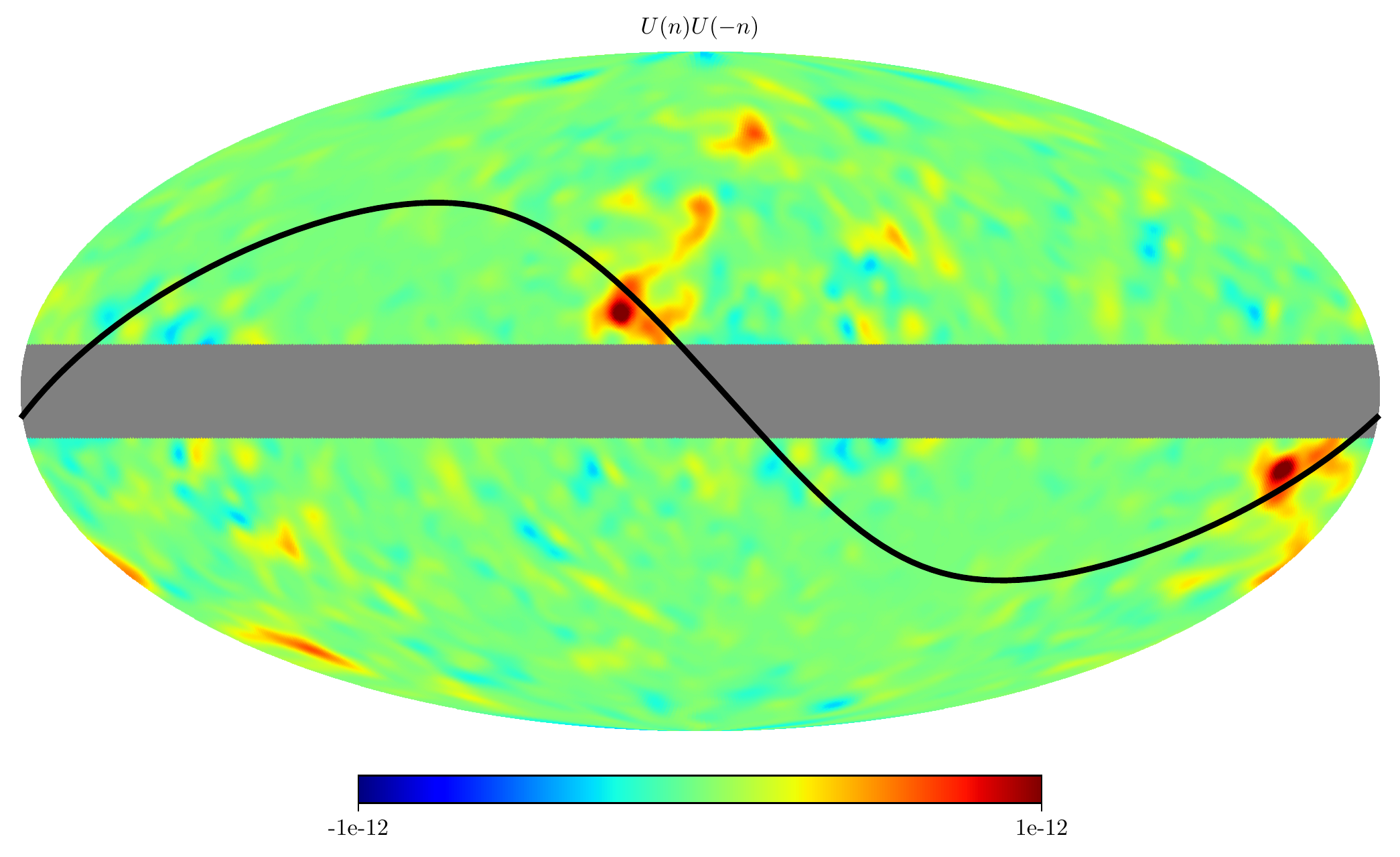}
    \caption{Asymmetry maps like $Z(\mathbf{n})$, but for the Stokes parameters $Q$ (left) and $U$ (right). A belt mask is applied.}
    \label{fig:10}
\end{figure*}

Finally, we note that the analysis of the anomalies in the polarization domain has important potential as an independent data set for verification of significance.
Determination of whether or not the anomalies can also be detected in the polarization data is a current research question.
We would like to point out that the methods used in this paper can be applied likewise to the polarization.
In figure~\ref{fig:10}, the asymmetry maps for the Stokes parameters $Q$ and $U$ are shown, computed as
\begin{align}
    Z_Q(\mathbf{n}) &= Q(\mathbf{n}) Q(-\mathbf{n}); \\
    Z_U(\mathbf{n}) &= U(\mathbf{n}) U(-\mathbf{n}). 
\end{align}
Especially in the $Z_U$ map, there is a prominent pair of positive-symmetry peaks, which is almost perfectly aligned with the RA, near to the pair of peaks 1a/1b in temperature (compare figure~\ref{fig1}) and the dipole modulation direction.
This zone indicates the possibility that the polarization may be involved in the low-multipole anomalies.

\section{Acknowledgement}
The HEALPix pixelization scheme was used in this work\cite{Gorski_2005}. This research was partially funded by Villum Fonden through the Deep Space project.

\bibliography{references}{}

\end{document}